\documentclass[12pt,a4paper]{article}
\usepackage{times}
\usepackage{dfttrob}
\usepackage[numbers,sort&compress]{natbib}
\usepackage{latexuseful2e}
\usepackage{amsmath}
\usepackage{graphicx}

\dfttnum{DFTT 22/2002\\July 2002}

\begin{document}
\newcommand{\glog}{g_{\text{log}}}

\title{FORWARD-BACKWARD MULTIPLICITY CORRELATIONS AND
LEAKAGE PARAMETER BEHAVIOUR IN ASYMMETRIC HIGH ENERGY COLLISIONS}
\author{A. Giovannini and R. Ugoccioni\\
 \it Dipartimento di Fisica Teorica and I.N.F.N - Sez. di Torino\\
 \it Via P. Giuria 1, 10125 Torino, Italy}
\maketitle

\begin{abstract}
Continuing previous work,
forward-backward multiplicity correlations are studied in asymmetric
collisions in the framework of the weighted superposition mechanism of
different classes of events.
New parameters for the asymmetric clan distribution and for the
particle leakage from clans in one hemisphere to the opposite one are
introduced to effectively classify different classes of collisions.
This tool should be used to explore forward-backward multiplicity
correlations in AB and pA collisions in present and future experiments
at RHIC and LHC.
\end{abstract}

\section{Introduction: the weighted superposition
mechanism of different classes of events}
It has been found quite recently \cite{RU:FB} that 
the weighted superposition of different classes of events  with  
negative binomial (NB) properties (events with and without mini-jets in
$pp$ collisions, 2-, 3- or more-jets samples of events in \ee\
annihilation)
reproduces,  in the GeV energy range, the available
experimental data on forward-backward (FB) multiplicity correlations of
the two types of collisions;
it is an intriguing result which sheds new
light on long range properties in multiparticle dynamics
and outlines the experimental consistency of the weighted
superposition mechanism of different classes of events.
In addition, the same  superposition mechanism 
has been shown  to provide in the GeV region an interesting phenomenological
tool in order to describe the observed shoulder effect in $n$-charged particle 
multiplicity distributions (MD's)
and $n$-oscillations in the ratios of  $n$-particle 
factorial moments  to the $n$-particle factorial cumulant 
moments \cite{DEL:1,hqlett+hqlett:2+frascati97:RU,L3:mangeol}. 
Effects on which QCD has not (up to now)
satisfactory predictions \cite{L3:mangeol}.

In the case of $pp$ collisions,
essential part of the theoretical background in the GeV region has been the 
assumption that, in each substructure or component described by a 
negative binomial (Pascal) multiplicity distribution [NB(P)MD], 
independently produced clans (they follow a Poisson distribution) are 
binomially distributed in the two hemispheres  and that logarithmically 
produced charged particles within each clan distribute themselves  again 
binomially in the two hemispheres  with an energy independent `leakage 
parameter.'
This parameter controls the number of particles,  generated by 
clans in one hemisphere, falling in the opposite one and was determined 
in \cite{RU:FB}  from the data at 63 GeV and 900 GeV.
Accordingly,  the correct reproduction in the GeV region  of the 
experimentally observed increase with c.m.\ energy of the forward-backward 
multiplicity correlation strength as well as of the relation between  
the average charged particle multiplicity of  particles lying  in the
backward hemisphere  versus the number of charged particles lying in the 
forward hemisphere (and vice versa)  support strongly our approach.

In the case of \ee\ annihilations,
in addition, it should be pointed out that under the same assumptions
of binomial distributions of clans and of particles generated by clans
in the forward and backward hemispheres, the relatively small value of
the forward-backward correlation strength in the total charged
particle MD and the absence of FB multiplicity
correlations in the separate 2- and 3-jet samples of events,
measured by OPAL collaboration \cite{OPAL:FB}, have been also correctly
reproduced.  Forward-backward correlations in the total charged
particle multiplicity distribution are indeed only due here to the
superposition effect of the two samples of events.
In our approach, in fact, forward-backward multiplicity
correlations in the two separate sample of events, each of NB type,
turn out to be zero, as no leakage is found from one hemisphere to
the other in the two separate samples, and in the total sample
resulting from the weighted superposition of two NB(P)MD we predict a
correlation which coincides with the experimental one within
experimental error \cite{RU:FB}.

The striking difference in FB charged particle multiplicity
correlation strength for the total sample of events  between proton
proton collisions in the GeV energy domain and  \ee\  annihilation 
at LEP energy  outlines the deep link between FB multiplicity correlations and 
long range correlations, which are expected to be quite strong in the
first case and relatively weak in the second one. In addition,   the observed
lack of FB multiplicity correlations  in the two separate  2-jet and 3-jet 
samples of events in \ee\  suggests that the weak FB correlations
seen in this reaction in the total sample of events are entirely due to
the superposition of  the two separate samples. This fact can be considered 
indeed an experimental evidence of the weighted superposition mechanism
characteristic of our approach (justified up to now only on the basis
of a quite successful description of collective effects  in high energy
energy phenomenology which QCD is unable at present to describe) and 
of the presence or absence of charged particle leakage from one hemisphere
to the opposite one.

In terms of clan structure analysis of the  NB behaviour for each class of
events, it is clear that  charged particle leakage as well as FB 
multiplicity and long range correlations are expected to be stronger when
particle population per clan is larger, a phenomenon which usually goes
together with the reduction of the average number  of clans.

It should be pointed out  that the occurrence of a larger average number
of smaller size clans and of a smaller average number  of larger size
clans has  a suggestive interpretation at parton level,
in the framework of a two-step mechanism,
and could be related  to  smaller  and larger  colour flow densities
respectively. In a NB description of final charged particle MD for a single
component (soft or semi-hard in $pp$ collisions, 2- and 3-jet samples of
events in  \ee\  annihilation), after using generalised local parton
hadron duality (GLPHD), the first step of the parton production process  
is dominated by the $A_{q\to q + g}(N_{\text{color}},\epsilon)$ vertex and the
average number of partonic clans corresponds to  the average number of
bremsstrahlung gluon jets (BGJ), which are effective independent intermediate
gluon  sources (IIGS), and the second step is controlled by the vertex 
$A_{g\to g+g}(N_{\text{color}},\epsilon)$, the gluon self-interaction 
vertex, whose 
increase  corresponds to  an enhancement of parton cascading from the
IIGS \cite{AGQCD,AGLVH:3+Bregenz}. 
$N_{\text{color}}$ is the number of colours and
$\epsilon$ the fixed cut-off regularization prescription of the theory.

Coming to the difference between  \ee\  and $pp$ reactions, it is likely
to expect that the observed increase of the average number of clans 
in \ee\  with respect to the $pp$ case  is a consequence
of a stronger activity  of the first vertex  with respect to the second one,
and that the opposite  will occur for the increase of the average number
of partons per partonic clans. It is quite clear  that the production
of a large
average number of partons per clan is a consequence of  a
longer cascading process,  originated  by IIGS generated  at relatively
high virtuality  in regions where, being the coupling constant smaller,
stronger colour flow between partons should be at work (a situation
favoured in $pp$ collisions). When the average number of partons per clan 
is relatively small, IIGS are expected to start to be effective at
lower virtuality, their cascading becomes shorter and colour exchanges
reduced with respect to the previous case (a situation favoured in
\ee\ annihilation). It seems therefore that the occurrence of stronger
FB multiplicity correlations, long range correlations and related
particle leakage enhancement from one hemisphere to the opposite one
at hadron level are a specular image of stronger cascading from high
virtuality IIGS and of larger colour exchange in this region at parton
level. 

We believe in fact that the understanding of FB multiplicity
correlations at hadron level is a possible starting point in order to
study new effects of colour quantum number exchanges in multiparticle
dynamics.

Accordingly, we decided to continue our search initiated in Ref.~\cite{RU:FB},
where FB correlations have been understood in the GeV region 
in symmetric reactions (like $pp$ collisions and \ee\ annihilation)
and  for symmetric  definition of the hemispheres by assuming  
at hadron level:

\begin{itemize}
\item[a.] NB behaviour for the (forward plus backward) MD of each component or
substructure (class  of events), i.e., clan structure analysis is 
assumed to be applicable to each of them. The generalisation to the
class of compound Poisson distributions
(CPD) is of course possible: the generating function
$G_{\text{CPD}}(z)$ in this case can be written as follows
\begin{equation}
	G_{\text{CPD}}(z) = \exp\left\{ \Nbar_g [g_c(z)-1 ] \right\}
\end{equation}
where $\Nbar_g$ is the average number of generalised clans and
$g_c(z)$ the generating function of the MD of charged particles
originated by a generalised clan (e.g., it is a logarithmic
distribution when the MD for a single component is a NB(P)MD).

\item[b.] clans are therefore independently emitted and their distribution is
Poissonian. In case of a generic CPD we talk, as previously stated, of
generalised clans.

\item[c.] clans are binomially distributed in the forward and backward
hemispheres (as are generalised clans).

\item[d.] logarithmically produced charged particles in each clan are also 
binomially distributed in the forward and backward hemispheres but
with different probabilities $p$ and $q$ ($p+q=1$, $p$ different from $q$);
$p$ controls the leakage from one hemisphere to the other: $p=1$
means that no particle leaks, while $p$ larger or equal to 0.5 and
smaller than 1 indicates leakage. 
In case of a generic CPD, the $n$-charged particle MD is generated by
$g_c(z)$, but $p$ and $q$ retain their meaning as described.

\end{itemize}

In order to perform our calculations in  $pp$ collisions  one extra assumption
has been added, i.e., that the  particle leakage parameter is constant
throughout the GeV region for the two separate  classes of events.

Possible scenarios in the TeV region based on extrapolations from data in the
GeV region and on the  weighted superposition mechanism of soft (without
mini-jets) and semi-hard (with mini-jets) events  have been indeed studied
and predictions given on $n$-charged particle multiplicity distributions and
on the ratios of $n$-particle factorial moments  to the $n$-particle factorial
cumulants moments  general properties in the TeV region
\cite{combo:prd+combo:eta}.

An attempt
to predict the energy dependence of the  forward-backward multiplicity
correlation strength  as well as $\nbar_B(n_F)$ vs $n_F$ 
general trends in the new energy range available at CERN  with Alice
detector has been considered
\cite{RU:FB}. The word is now to experiment which is
supposed to test all these three sets of predictions at 14 TeV.

The problem we want to face in this paper, 
on the theory side, is the generalisation  of the
approach discussed in Ref.~\cite{RU:FB} 
for determining forward-backward multiplicity 
correlations  properties for a single component,
in more complex asymmetric reactions (like heavy 
ion  AB and $p$A collisions) and to provide a general framework for
the study of forward-backward multiplicity correlations  which includes
symmetric reactions (like $pp$  and
AA collisions and \ee\  annihilation) as a particular case.

Corner stones of our argument remain of course assumptions  a, b, c, d.

They could hardly be abandoned in view of their success in giving a good 
phenomenological description of available experimental data on  FB 
multiplicity correlations  in symmetric reactions.

The  generalisation of the approach to asymmetric reactions  will concern 
therefore    mainly  how to implement,  in the framework defined 
by assumptions a.b.c.d, the asymmetry of the reaction and the asymmetric
definition of the forward and backward hemispheres.

In view of the lack of sound experimental data on asymmetric reactions
and of the related analyses in terms of two or more component
substructures both in full phase space and in rapidity intervals, the
present paper should be considered as a stimulus to experimentalists
of the new generation machines for a deeper analysis of total
$n$-charged particles MD and a more satisfactory understanding of
forward-backward multiplicity correlations as the c.m.\ energy
increases and at fixed c.m.\ energy within rapidity intervals.

\section{The general asymmetric case}

Generalisation of the study performed in Ref.~\cite{RU:FB} on FB 
multiplicity correlations  consists at the present stage 
of investigation  in  assuming that

\begin{itemize}
\item[A.] the particle leakage percentage from the backward (B) 
to the forward (F) hemisphere, $q_B$, is different 
from the particle leakage percentage  from F to B hemisphere, $q_F$, with
$p_F + q_F = 1$ and  $p_B +q_B =1$, $p_F$ and $p_B$ being the corresponding
percentages of particles not leaking from one hemisphere to the opposite
one and remaining in the F and B hemisphere respectively; 
notice that in Ref.~\cite{RU:FB}, $p_F=p_B=p$.

\item[B.] Poissonianly generated clans are asymmetrically 
(but binomially) distributed in the
two hemispheres  with the asymmetry parameter $r$ different from 1/2, $s$ 
being its complement to 1 in the opposite hemisphere; in
Ref.~\cite{RU:FB},  notice that $r=s=1/2$ in addition to $p_F=p_B=p$.

\end{itemize}

\subsection{The generating function}

In general, the joint MD $P(n_F,n_B)$
for $n_F$ particles in the forward hemisphere
and $n_B$ particles in the backward one is related to the 
global MD $P(n)$ though the probability distribution $f(n_F|n)$, which
gives the probability to have $n_F$ F-particles when the total
number of particles is $n$, as follows:
\begin{equation}
		P(n_F,n_B) = P(n_F + n_B) f(n_F | n_F + n_B) .
\end{equation}
The generating function $G(z_F,z_B)$ for the joint distribution then
satisfies
\begin{equation}
		G(z_F,z_B) \equiv \sum_{n_F,n_B} z_F^{n_F} z_B^{n_B} P(n_F,n_B)
	   = \sum_{n} z_B^n P(n) g_f(z_F/z_B; n) ,
				\label{eq:2}
\end{equation}
where $g_f(z; n)$ is the generating function for $f(n_F | n)$;
in case it is the binomial distribution with parameter $p$,
\begin{equation}
	f(n_F | n) = \binom{n}{n_F} p^{n_F} (1-p)^{n- n_F},
\end{equation}
then its GF is (defining $q \equiv 1-p$):
\begin{equation}
	g_f(z; n) = (q + p z)^n  ,
\end{equation}
and one obtains a considerable simplification of Eq.~(\ref{eq:2}):
\begin{equation}
	G(z_F,z_B) = g( z_F p  +  z_B q ),
\end{equation}
where $g(z)$ is the generating function of the global distribution
$P(n)$:
\begin{equation}
	g(z) \equiv \sum_{n} z^n P(n) .
\end{equation}

Accordingly, we proceed now to calculate the GF for the
general case. The formulae below are heavily based on previous
work \cite{RU:FB}.  We consider a clan with logarithmic MD
produced in the F hemisphere and assume that each particle
has the same probability $p_F$ not to leak into the B hemisphere.
Then the distribution at fixed number
of particles is binomial, and we have immediately an application
of the just-explained scheme: 
the GF for the joint distribution within a F-clan is thus
\begin{equation}
	g_{c,F}(z_F,z_B) = \glog(z_F p_F  +  z_B q_F) ;
																		\label{eq:gcF}
\end{equation}
for a clan produced in the B hemisphere we have the corresponding
GF:
\begin{equation}
	g_{c,B}(z_F,z_B) = \glog(z_F q_B  +  z_B p_B) .
																		\label{eq:gcB}
\end{equation}
Here the GF for the logarithmic distribution of parameter $\beta$ is 
\begin{equation}
	 \glog(z) \equiv \log(1-z \beta)/\log(1-\beta) ,
\end{equation}
where, in terms of standard NB parameters $\nbar$ and $k$,
\begin{equation}
	\beta = \frac{\nbar}{\nbar + k}
\end{equation}
and is related to the average number of particles per clan
via the formula:
\begin{equation}
	\nc = \frac{\beta}{(\beta-1)\log(1-\beta)} .
\end{equation}

Recalling now Eq.~(34) 
of \cite{RU:FB}, it is easy to calculate the GF
of the joint distribution of F and B particles at given numbers $N_F$
of F clans and $N_B$ of B clans:
\begin{equation}
		g(z_F,z_B|N_F, N_B) = \left[ g_{c,F}(z_F,z_B) \right]^{N_F}
					\left[ g_{c,B}(z_F,z_B) \right]^{N_B} :
\end{equation}
due to the fact that all clans are by definition independent from each
other, we can convolute the respective MD's, which corresponds to
multiply together the GF's.

In order to sum over the clan MD, let us remember that we have assumed
that clans are Poisson distributed and independent of each other,
thus the joint distribution can be written as:
\begin{equation}
	{\cal P}(N_F,N_B) = \frac{\Nbar^{N_F+N_B}}{(N_F+N_B)!}
			e^{-\Nbar} \binom{N_F+N_B}{N_F} r^{N_F} (1-r)^{N_B},
\end{equation}
where $\Nbar = \Nbar_F + \Nbar_B$ is the average number of clans, and
$r = \Nbar_F/\Nbar$ is the fraction of clans emitted in the F hemisphere. The
corresponding GF is
\begin{equation}
	{\cal G}(z_F,z_B) = \exp\left\{ \Nbar [r z_F + (1-r) z_B - 1] \right\}.
\end{equation}

We can now perform the last step in the calculation, again exploiting
the general properties of the binomial distribution:
\begin{equation}
	g(z_F,z_B) = \sum_{N_F}\sum_{N_B} g(z_F,z_B|N_F, N_B) 
						{\cal	P}(N_F,N_B) =
		{\cal G} \left( g_{c,F}(z_F,z_B), g_{c,B}(z_F,z_B) \right).
																			\label{eq:14}
\end{equation}
\begin{equation}
	g(z_F,z_B) = \exp \left\{ 
				r\Nbar [ \glog(z_F p_F + z_B q_F) - 1 ]
			\right\} \exp \left\{
				(1-r)\Nbar [ \glog(z_F q_B + z_B p_B) - 1 ]
			\right\}   .
    \label{eq:1}
\end{equation}
Notice that the above formula is valid for the NB(P)MD, but
can easily be extended, as anticipated,
to any compound Poisson distribution, provided all correlations
are exhausted within a (generalised) clan: it is sufficient to
replace $\glog$ with the appropriate GF within a clan, $g_c(z)$.

\subsection{The correlation strength and the marginal distributions}

We can now use the above result to deduce the correlation strength
$b$:
\begin{equation}
	b = \frac{\avg{n_F n_B} - \nbar_F \nbar_B}
			{\left[ (\avg{n_F^2}-\nbar_F^2) (\avg{n_B^2}-\nbar_B^2)
			\right]^{1/2}},
													\label{eq:b_defined}
\end{equation}
since everything can be read from the GF:
\begin{equation}
	\avg{n_F n_B} = \left. \frac{\partial^2 g}{\partial z_F\partial z_B}
			\right\vert_{z_F=z_B=1}
\end{equation}
\begin{equation}
  \nbar_i = \left. \frac{\partial g}{\partial z_i}
			\right\vert_{z_F=z_B=1} 
	\qquad
	\avg{n^2_i} = \left. \frac{\partial^2 g}{\partial z_i^2}
			\right\vert_{z_F=z_B=1} + \nbar_i
\end{equation}
with $i=$ F, B.

The average number of F-particles at fixed number of B-particles
is also easily obtained though differentiation, since the MD
of $n_F$ at fixed $n_B$ is given by
\begin{equation}
  p(n_F|n_B) = \frac{ p(n_F,n_B) }{ p_{\text{marg}}^{(B)}(n_B)}
			= \frac{ p(n_F,n_B) }{ \sum_{n_F} p(n_F,n_B) },
\end{equation}
where $p_{\text{marg}}^{(B)}(n_B)$ is the marginal distribution
in the B hemisphere;   the GF is
\begin{equation}
	g(z_F|n_B) \equiv \sum_{n_F} z_F^{n_F} p(n_F|n_B)
						= \left. \frac{\dfrac{\partial^{n_B}}{
														\partial z_B^{n_B}} g(z_F,z_B)}{
						\dfrac{\partial^{n_B}}{
														\partial z_B^{n_B}} g(1,z_B)
		}\right\vert_{z_B=0}  .
\end{equation}
being $g(1,z_B)$ the GF of $p_{\text{marg}}^{(B)}$.
The first moment is obtained from the GF by
differentiating once:
\begin{equation}
	\nbar_F(n_B) \equiv \left.\frac{\partial g(z_F|n_B)}{\partial z_F}
			\right\vert_{z_F=1} =
			\left. \frac{\dfrac{\partial^{n_B}}{
														\partial z_B^{n_B}} 
							\dfrac{\partial}{\partial z_F} g(z_F,z_B)}{
						\dfrac{\partial^{n_B}}{
														\partial z_B^{n_B}} g(z_F,z_B)
		}\right\vert_{z_B=0,\,z_F=1}  .
														\label{eq:avgnF}
\end{equation}

Analogous formulae for the B hemisphere quantities can be easily
obtained in the same way.

The corresponding marginal distributions 
are $g(z_B = z, 1)$ and $g(1 , z_F =z)$:
\begin{equation}
g(z,1)= \exp \left\{ r \Nbar [ \glog (z p_F + q_F) - 1 ] \right\}
        \exp \left\{ s \Nbar [ \glog (z q_B + p_B) - 1 ] \right\} ;
				\label{eq:Al2}
\end{equation}
$g(1,z)$ can be obtained from Eq.~(\ref{eq:Al2})
by interchanging parameter $p_i$ with $q_i$ (i=F,B).

The  marginal distribution  of Eq.~(\ref{eq:Al2}) is the product of 
the  GF's of two NB(P)MD's  with characteristic NB 
parameters \{$\nbar r p_F$ , $k r$\} and
\{$\nbar s q_B$, $k s$\} respectively as can be seen immediately by noticing
that  
\begin{equation}
g(z,1) =  \left\{ 1 + \frac{\nbar r p_F}{k r} ( 1 - z )\right\}^{ - kr}
          \left\{ 1 + \frac{\nbar s q_B}{k s} ( 1 - z )\right\}^{ -
          ks} ;
\end{equation}
$g(1,z)$ can be defined in an analogous way.

$g(z,1)$ and $g(1,z)$ are the GF's of the MD's obtained by convoluting the MD's
for particles generated by F (B) clans  which  stay in the F (B) hemisphere  
and  do not leak in the opposite hemisphere with the MD's for particles  
generated by B (F) clans  which  are leaking in the F (B) hemisphere.

Although $g(z,1)$  (and $g(1,z)$) is the product of the GF's of NB(P)MD's it
is not the GF of a NB(P)MD. This consideration notwithstanding, it is 
interesting to remark that $g(z,1)$ (and $g(1,z)$) is  an infinitely
divisible distribution (IDD),  a fact which allows to define the
generalised clan concept, as already remarked.

In fact  Eq.~(\ref{eq:Al2})  can be rewritten as follows
\begin{equation}
g(z,1)= \exp \left\{ \Nbar [r \glog (z p_F + q_F) + 
		s \glog (z q_B + p_B )] - \Nbar \right\}
\end{equation}
i.e., as
\begin{equation}
   g(z,1) = \exp \left[ \Nbar ( A(z) - 1 ) \right]
				\label{eq:Al3}
\end{equation}
with $A(z)= [r  \glog (z p_F + q_F) + s  \glog (z q_B + p_B )]$
and $A(0)$ different from zero.
This last remark implies that the probability of generating zero 
particles is different from $e^{- \Nbar}$. Therefore the probability
of generating empty clans is also different from zero, a result
in contrast with  the standard clan definition (each clan  contains at  
least one particle), definition which we  would like of course to enforce.

In order to do that,  let us add and subtract in the exponent of
Eq.~(\ref{eq:Al3})  the term $A(0)$, i.e., we rewrite
Eq.~(\ref{eq:Al3}) as follows
\begin{equation}
g(z,1) = \exp \left\{ \Nbar [ A(z) - A(0) + A(0) - 1 ] \right\}  .
				\label{eq:Al4}
\end{equation}
From Eq.~(\ref{eq:Al4}) one obtains  the compound Poisson 
distribution  belonging to the class of IDD
\begin{equation}
g(z,1) = \exp \left\{ \Nbar_g [ G_g  (z) - 1 ] \right\}  ,
				\label{eq:Al5}
\end{equation}
with $\Nbar_g = [1 - A(0)]\Nbar$ and  $G_g(z) = [A(z)- A(0)] [ 1 -
A(0)]^{-1}$. 

Accordingly, one can define   the average number of generalised
clans, $\Nbar_g$,  in terms of the standard  NB parameters 
$\nbar$ and $k$, and of $p_F$ and $p_B$:
\begin{equation}
\begin{split}
\Nbar_g &= \Nbar [1 - A(0)] = \Nbar \{1 - [r \glog (q_F) + s \glog
                              (p_B)]\} \\
              &= - \ln g(0,1)
            = { r k \ln ( \nbar +  k - \nbar q_F) +
                s k \ln ( \nbar +  k - \nbar p_B) - k \ln k } .
			\label{eq:Al6}
\end{split}
\end{equation}
The backward marginal multiplicity distribution and related properties can be
be obtained  from the forward one
by interchanging parameter $p_i$ with $q_i$ (i=F,B).

\section{The cases of partially removed  symmetry}

\subsection{Symmetry for clans only ($r = s = 1/2$, $p_F \ne p_B$,
$q_F \ne q_B$)}   
 
The symmetry of the reaction can be broken partially by assuming that
binomially distributed clans go fifty per cent in the F hemisphere and
fifty per cent in the B hemisphere ($r = s = 1/2$) but particle
leakage from clans in F to B hemisphere, $q_F$, and particle leakage
from clans in B to F hemisphere, $q_B$, (as their complements $p_B$
and $p_F$ with $p_F+q_F = 1$ and $p_B + q_B = 1$) are different.
In other words, assumption B from the previous section
is valid but A is not.

The forward-backward joint charged  particle multiplicity GF
for one single component  turns out to be 
\begin{equation}
g(z_B,z_F) = \exp \left\{ \tfrac{1}{2}\Nbar 
		[ \glog(z_F p_F + z_B q_F) - 1 ] \right\}
              \exp \left\{ \tfrac{1}{2}\Nbar 
    [ \glog(z_B p_B + z_F q_B) - 1 ]\right\} ,
				\label{eq:Al7}
\end{equation}
and for the corresponding  marginal B  distribution ($z_B = z$, $z_F=1$)
\begin{equation}
g(z,1) = \exp \left\{ \tfrac{1}{2}\Nbar [ \glog( p_F + z q_F) - 1 ]  \right\}
       \exp \left\{ \tfrac{1}{2}\Nbar [ \glog(z p_B +  q_B) - 1 ]
       \right\} .
				\label{eq:Al8}
\end{equation}
In terms of standard NB parameters Eq.~(\ref{eq:Al8})  becomes
\begin{equation}
\begin{split}
g(z,1) &= \left[ 1 + \frac{\nbar p_F}{k}  (1-z) \right]^{- k/2 }
        \left[ 1 + \frac{\nbar q_B}{k}   (1-z) \right]^{- k/2}\\
       &= \left[ 1 + \frac{\nbar}{k}  (1-z)  + 
								\frac{\nbar^2}{k ^2}  p_F q_B (1- z) 
				\right]^{-k/2} .
	\label{eq:Al9}
\end{split}
\end{equation}

The  symmetric definition of the two hemispheres is 
removed here  in one component by assuming that  particle  leakage parameters
in the forward and backward hemispheres are different.

The GF  of the forward marginal multiplicity distribution  thanks to the
quadratic term  as already shown in the general case  is no more a NB(P)MD, 
although it is the product of the GF's  of two NB(P)MD's with characteristic 
parameters
\{$\frac{1}{2}  \nbar p_F$, $\frac{1}{2} k$\}  and 
\{$\frac{1}{2}  \nbar q_B$, $\frac{1}{2} k$\}  with asymmetric
average charged particle multiplicities.
The GF of the marginal distribution
becomes the GF of a NB(P)MD in some special cases,  i.e., for  $q_B = 0$, 
corresponding  to no particle leakage from the  backward  to the forward
hemisphere, and for $\nbar \ll k$,  
a situation  which occurs in the Poissonian
limit for average charged particle multiplicity much less than the $k$ 
parameter
and almost coinciding with the average number of clans. The GF $g(z,1)$,
although not  NB,  is still  an infinitely divisible distribution,
as expected.

\subsection{Symmetry for particles within clans only ($r \ne s \ne
1/2$, $p_F=p_B$, $q_F=q_B$)}   

Another way to remove partially the symmetry is to 
use assumption A without B, i.e., to assume that
binomially distributed clans are not symmetrically subdivided between 
F and B hemispheres, but particle leakage from clans in one hemisphere to the
other is the same ($r \ne s \ne 1/2$  with $r+s=1$,  and
$p_F=p_B= p$, $q_F=q_B=q$  with $p+q=1$, and $p$ larger or equal than 1/2  and
smaller than 1). 

The FB joint particle multiplicity distribution GF for
one component, $g(z_F,z_B)$, becomes in this case
\begin{equation}
g(z_F,z_B) = \exp \left\{ \Nbar r [ \glog(z_F p + z_B q )- 1 ] \right\} 
              \exp\left\{ \Nbar s [ \glog(z_B p + z_F q )- 1 ]
              \right\} ,
			\label{eq:Al10}
\end{equation}
and the  corresponding forward  marginal charged particle MD GF
\begin{equation}
g(z,1)= \exp \left\{ \Nbar r[ \glog(z p +  q) - 1 ] \right\} 
              \exp \left\{ \Nbar s[ \glog(z q+ p)- 1 ] \right\} ,
			\label{eq:Al11}
\end{equation}
which in terms of standard NB parameters of the total charged particle MD
of the component under investigation is
\begin{equation}
 g(z,1)=\left[1+ \frac{\nbar p}{k} (1-z) \right]^{- k r}
        \left[ 1 + \frac{\nbar q}{k}  (1-z) \right]^{- k s} .
			\label{eq:Al12}
\end{equation}

The forward marginal MD GF, although again not of NB type, is the
product of the GF's of two NB(P)MD's with parameters \{$\nbar r p$, $k r$\} and
\{$\nbar q s$,  $k s$\} respectively. 
The  two sets of parameters characterise
the group of  non-leaking particles lying in the F hemisphere and those
which leak from the backward hemisphere to the forward one.  The backward
marginal distribution can be easily obtained from the above equations 
by interchanging parameter $r$ with $s$.

\section{The symmetric case ($r=s=1/2$, $p_F=p_B$, $q_F=q_B$)}

In this case, assumptions A and B are both rejected.
The formula studied in \cite{RU:FB} for  the joint charged particle MD GF
in the symmetric case follows from Eq.~(\ref{eq:1}) by taking  $p_F=p_B=p$,
$q_F=q_B=q$ with $p$ different from $q$, $p+q=1$, and $r=s=1/2$.

We get
\begin{equation}
\begin{split}
g(z_B,z_F) &= \exp \left\{\tfrac{1}{2}\Nbar  [ \glog (z_F p + z_B q) - 1] 
                            [\glog(z_B p + z_F q) - 1] \right\} \\
           &= k \frac{[k+\nbar (1-pz_B-qz_F)]^{k/2}}{
								 [ k + \nbar( 1-p z_F -q z_B)]^{k/2} }
			\label{eq:Al13}
\end{split}
\end{equation}
and for the corresponding forward marginal distribution $g(z,1)$
\begin{equation}
\begin{split}
g(z,1) &= [1 + \frac{\nbar q}{k}(1-z)]^{- k/2} 
          [1 + \frac{\nbar p}{k}(1-z)]^{- k/2} \\
       &= \left\{1 + \frac{\nbar}{k}(1-z) + pq 
									\left[ \frac{\nbar}{k} (1-z)\right]^2 \right\}^{-
          k/2} .
			\label{eq:Al14}
\end{split}
\end{equation}
In conclusion, Eq.~(\ref{eq:1}) can be considered the wanted
generalisation of Eq.~(\ref{eq:Al13}) to the case of asymmetric
reactions and an asymmetric definition of the forward and backward
hemispheres.

\section{Behaviours of $\nbar_F(n_B)$ vs.\ $n_B$ and of
$\nbar_B(n_F)$ vs.\ $n_F$}

In this section we examine the relation between 
the average number of particles in one hemisphere,
$\nbar_F(n_B)$ or $\nbar_B(n_F)$, 
at fixed value of the number of particles, 
respectively $n_B$ or $n_F$,
in the other hemisphere, according to Eq.~(\ref{eq:avgnF}).

Let us start by recalling that the von Bahr-Ekspong theorem
\cite{Ekspong:theorem} implies,
since we assume the total MD to be of NB type,
that there is no linearity in $\nbar_B(n_F)$ vs $n_F$
(and in $\nbar_F(n_B)$ vs $n_B$)
unless the MD for F-particles (and for B-particles!)
at fixed total number of particles is binomial,
in which case the GF is simply:
\begin{equation}
	 \exp[ \glog(z_F p + z_B q) - 1 ] .
															\label{eq:Ekspong}
\end{equation}
Notice that it does not make sense to distinguish $p_F$ from $p_B$ in
this case; however, if one did not distinguish F-clans from B-clans,
i.e., if each clan emitted the same fraction $p$ in one hemisphere,
then in Eqs.~(\ref{eq:gcF}) and (\ref{eq:gcB}) above 
one would put $p_F = q_B = p$ and Eq.~(\ref{eq:1}) would reduce to
Eq.~(\ref{eq:Ekspong}). 

\begin{figure}
  \begin{center}
  \mbox{\includegraphics[width=\textwidth]{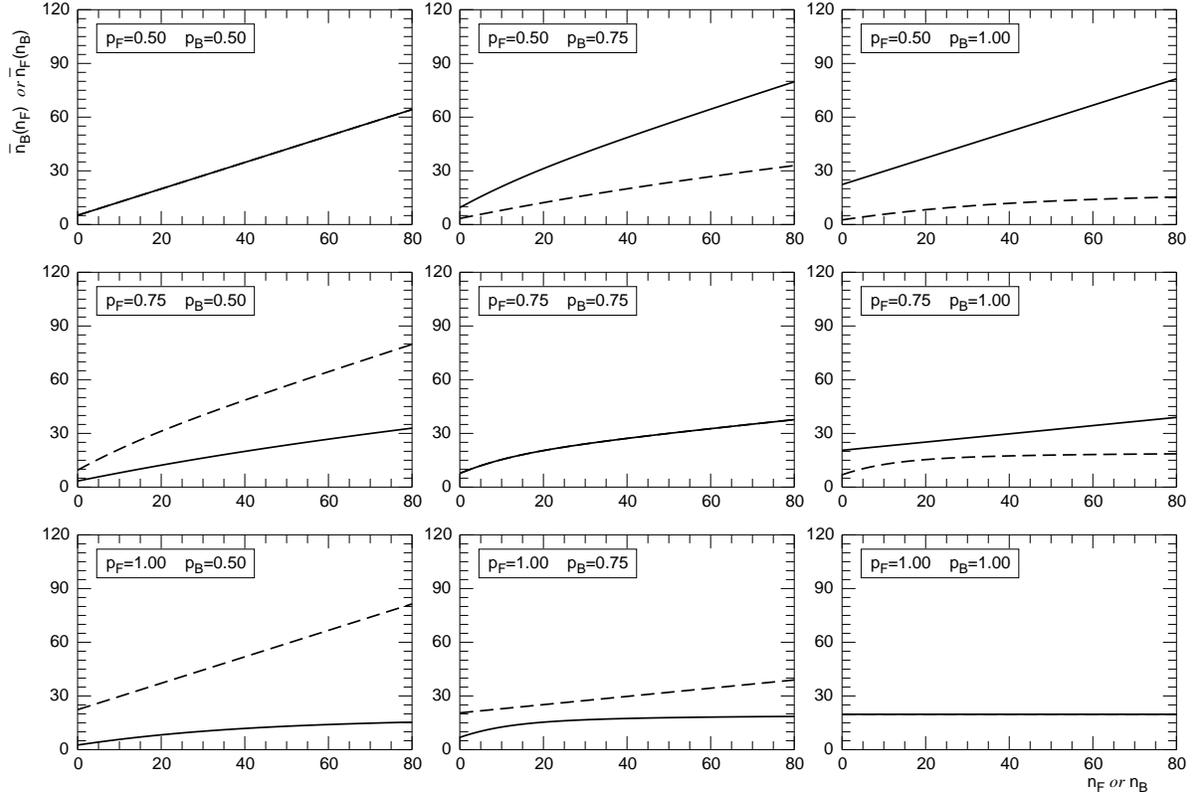}}
  \end{center}
  \caption{Behaviours of $\nbar_F(n_B)$ vs.\ $n_B$ (dashed lines) and of
$\nbar_B(n_F)$ vs.\ $n_F$ (solid lines) for $r=s=1/2$ and
values of the leakage parameters as indicated. In order to
illustrate a practical case, we have taken
for the NB parameters $\nbar=39.5$ and $k=7.0$, corresponding to the
soft component of $pp$ collisions at 14 TeV in the scenarios examined 
in Ref.~\cite{combo:prd+combo:eta}.
In panes where only the
solid line is visible, it means the dashed line coincides with it.
}\label{fig:r050}
  \end{figure}

\begin{figure}
  \begin{center}
  \mbox{\includegraphics[width=\textwidth]{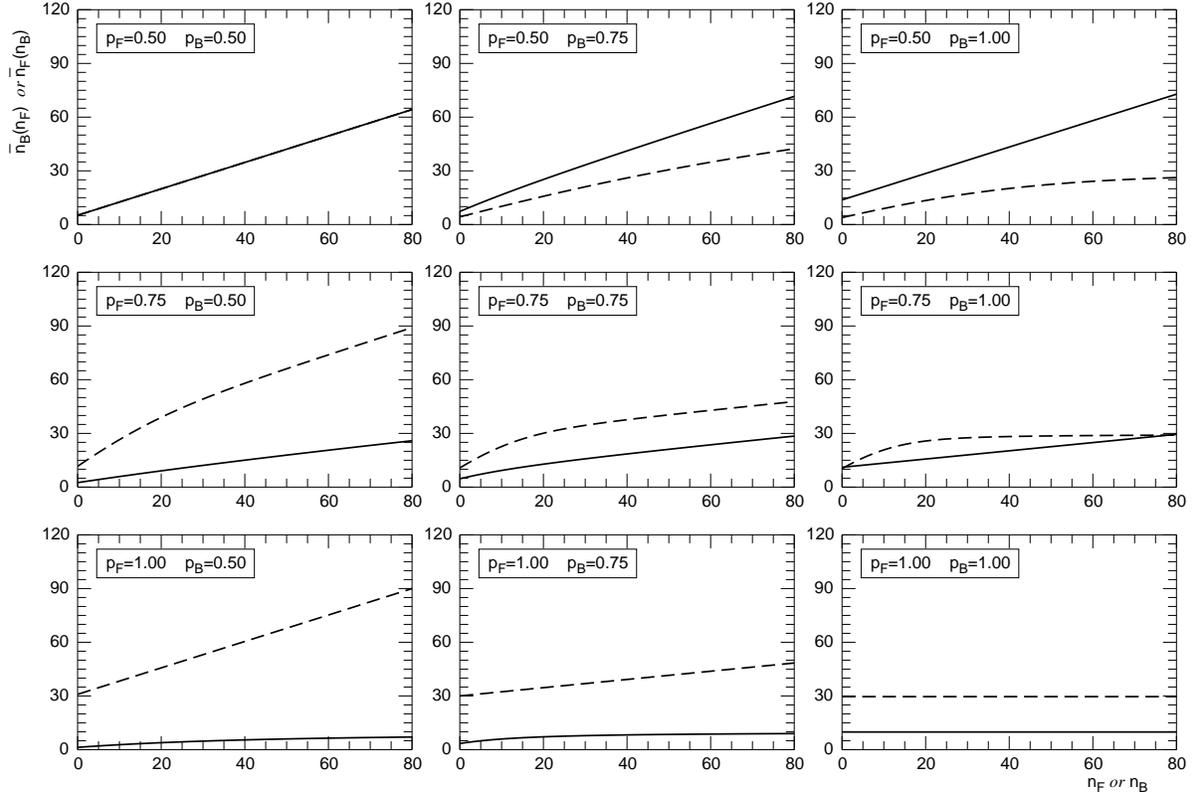}}
  \end{center}
  \caption{Same as Figure \ref{fig:r050} but for $r=3/4$, $s=1/4$.
}\label{fig:r075}
  \end{figure}

In the following
the symmetry in the clan distribution is contrasted with the
asymmetric clan distribution. In addition, leakage parameters for the
two components are taken either equal or different.
The resulting pictures of forward-backward multiplicity correlations
are shown in Fig.s~\ref{fig:r050} and \ref{fig:r075}: of course our
choice of the involved parameters is arbitrary. The word is again to
experiment on symmetric collisions ($r=s=1/2$) in order to understand
eventual deviations from identity of particle leakage parameters from
clans in the forward and backward hemispheres and to experiment on
asymmetric collisions ($r\ne s\ne 1/2$) in order to measure the
different leakage parameters in the two hemispheres. All together,
expected different values of $r$ and $s$ as well as $p_F$ and $p_B$
parameters could lead to a new intriguing classification, in terms of
FB multiplicity correlations, of high energy collisions and their
substructures.  

\subsection{The case $r=s=1/2$ (symmetry in the clan distribution)}

We start by examining the case in which $r = 1/2$,
illustrated in Figure~\ref{fig:r050}.
When $p_F = p_B = 1/2$, there is perfect linearity in $\nbar_B(n_F)$ 
versus $n_F$. This is in agreement with the mentioned theorem,
because here we are saying that, within each clan, particles are
binomially distributed in F and B with the same probability 1/2,
thus the fact that particles are produced in clans does not make any
difference. 
When $1/2 < p_F = p_B < 1$, on the contrary, the fact that the
production happens in two steps becomes again important and
the relation between $\nbar_B(n_F)$ and $n_F$ is non-linear.
Linearity is recovered again in the limiting case of no correlations
($p_F = p_B = 1$), see below.

When $p_F \ne p_B$, the symmetry is lost because the particles
leaking from the F hemisphere into B are not compensated (if $p_F < p_B$),
or are over-compensated (if $p_F > p_B$), by those leaking from B into F.
Thus one obviously finds that $\nbar_B(n_F)$ vs $n_F$ is not the same as 
$\nbar_F(n_B)$ vs $n_B$. Indeed the two 
marginal distributions differ from each other, but can be obtained
one from the other by exchanging $p_F$ with $p_B$.
In general for $p_F > p_B$ there is less leakage from F than from B:
particles prefer to stay in the F hemisphere thus
one has the line of $\nbar_F(n_B)$ vs $n_B$ above $\nbar_B(n_F)$ vs $n_F$.
Vice versa, the opposite is true for $p_F < p_B$.

One can notice that when $p_F = 1$, $\nbar_F(n_B)$ vs $n_B$ is a straight line,
and when $p_B = 1$, $\nbar_B(n_F)$ vs $n_F$ is linear. 
The fact is that when $p_F = 1$, clans that fall in the F hemisphere
do not contribute to the B hemisphere. The only relation between
$n_B$ and $\nbar_F$ is given by B clans leaking to the F hemisphere.
And this relation is by definition binomial for each clan: the
number of clans is Poisson distributed, thus the overall is
still binomial with the same parameter, as was shown above.
Indeed, the GF with $p_F=1$ is:
\begin{equation}
	g(z_F,z_B) = \exp \left\{ r\Nbar[ \glog(z_F) - 1]\right\} 
				\exp \left\{ (1-r)\Nbar
					[ \glog(z_F q_B + z_B p_B) - 1] \right\} ;
\end{equation}
applying Eq.~(\ref{eq:avgnF}), the first term gives just
the constant $r\nbar$, while the second one is of the same type
of Eq.~(\ref{eq:Ekspong}) and thus also gives a linear
behaviour. Notice that this happens independently of the value
of $r$.
A corresponding result can be obtained for $p_B = 1$ by
appropriately exchanging the roles of F and B.

\subsection{The case $r \ne s \ne 1/2$ (asymmetry in the clan distribution)}

When $r$ increases over 1/2
(e.g., in Figure~\ref{fig:r075} the case $r=3/4$ is illustrated), 
the excess of clans in the F hemisphere
implies an increase of $\nbar_F(n_B)$ vs $n_B$ and a decrease 
of $\nbar_B(n_F)$ vs $n_F$,
enlarging the differences between the curves when $p_F > p_B$
and reducing them when $p_F < p_B$ ($\nbar_F(n_B)$ vs $n_B$ 
can even become larger
than $\nbar_B(n_F)$ vs $n_F$, depending on the values of $r$, $p_F$ and $p_B$).
Exchanging $p_F$ with $p_B$ does not restore the symmetry.
If $p_F = p_B = 1/2$, then the value of $r$ has no influence on the outcome: 
if particles within each clan have a 50\% chance of going into the other
hemisphere, it makes no difference if clans prefer one hemisphere over
the other, and the two curves coincide again. 
This is not true if $p_F = p_B > 1/2$.
Finally, in agreement with the remark at the end of the last subsection,
we notice that even in the case $r > 1/2$ when $p_F = 1$,
$\nbar_F(n_B)$ vs $n_B$  
is a straight line, and when $p_B = 1$, $\nbar_B(n_F)$ vs $n_F$ is linear.

\subsection{Asymmetry in the average number of particles per clan}
All the above has been studied in the framework of a NB(P)MD
describing in one component the forward plus backward MD.
However, there are low energy data on pA collisions 
in which NB behaviour has been found to hold separately in the forward
and in the backward hemisphere, with different NB parameters \cite{Dengler}.
Because clans are independently emitted, the only correction to make
to our formulae is to allow for different average numbers 
of particles per clan in
the two hemispheres: in other words, the parameter $\beta$ of the
logarithmic distributions in Eq.s~(\ref{eq:gcF}) and (\ref{eq:gcB})
will be different, say $\beta_F$ and $\beta_B$. Now, we immediately
get 
\begin{equation}
	g(z_F,z_B) = \exp \left\{ 
				r\Nbar [ \glog(z_F p_F + z_B q_F; \beta_F) - 1 ]
			\right\} \exp \left\{
				s\Nbar [ \glog(z_F q_B + z_B p_B; \beta_B) - 1 ]
			\right\}
    \label{eq:1bis}
\end{equation}
in place of Eq.~(\ref{eq:1}). 
The product of these two NB(P)GF is not a
NB(P)MD, but it is still, by construction, an infinitely
divisible distribution.

\section{Conclusions}
It has been shown that assuming different particle leakage percentages
($p_B \ne p_F$) for binomially generated particles from clans in one
hemisphere to the opposite one and asymmetric ($r\ne s$) distribution
in the two hemispheres of binomially generated clans, a general
formula for the generating function of the joint 
$(n_F,n_B)$-charged particle
multiplicity distribution for each class of events (or
substructure) can be obtained when the total MD GF is of NB type. 
The formula reduces to that discussed in
Ref.~\cite{RU:FB} for $p_B = p_F$ and $r = s = 1/2$.
Of particular interest are also the cases in which the symmetry is
only partially removed (assumption A without B and B without A).
All above-mentioned results, although explicitly derived for
substructures of NB type, can be easily generalised to any discrete
infinitely divisible MD.
This search is relevant for the study of forward-backward
multiplicity correlations in non-identical heavy ion 
and in proton-nucleus collisions.
Accordingly, the newly introduced particle leakage and asymmetry
parameters can be considered as effective indeces
classifying different classes of collisions.



\end{document}